# Suppression of electron spin decoherence in Rabi oscillations induced by an inhomogeneous microwave field

A.P. Saiko[1], R. Fedaruk[2], and S.A. Markevich[1]

[1]Scientific-Practical Materials Research Centre NAS of Belarus, Minsk, Belarus

[2]Institute of Physics, University of Szczecin, 70-451, Szczecin, Poland

E-mail: saiko@ifttp.bas-net.by

**Abstract.** The decay of Rabi oscillations provides direct information about coherence of electron spins. When observed in EPR experiments, it is often shortened by spatial inhomogeneity of the microwave field amplitude in a bulk sample. In order to suppress this undesired loss of coherence, we propose an additional dressing of spin states by a weak longitudinal continuous radiofrequency field. The Gaussian, cosine and linear distributions of the microwave amplitude is analyzed. Our calculations of the Rabi oscillations between the doubly dressed spin states show that for all these distributions the maximum suppression of the inhomogeneity-induced decoherence is achieved at the so-called Rabi resonance when the radio-field frequency is in resonance with the Rabi frequency of spins in the microwave field. The manifestations of such suppression in the published EPR experiments with the bichromatic driving are discussed. The realization of the Rabi resonance using the radiofrequency field could open new possibilities for separating the contributions of relaxation mechanisms from those due to the inhomogeneous driving in spin decoherence.

Keywords: Pulsed EPR; Rabi oscillations; Spin decoherence; Inhomogeneous microwave field; Longitudinal radiofrequency field; Rabi resonance.

## 1. Introduction

Coherent manipulation of two-level spin systems (qubits) by electromagnetic fields is attracting interest in pulsed electron paramagnetic resonance (EPR) and in its many applications including quantum information technologies [1,2]. Paramagnetic ions and defects in diamagnetic solid matrixes are among the promising candidates for solid-state qubits [3-5]. Rabi oscillations [6] (or transient nutations [7]) represent the basic phenomenon used for coherent manipulation of spin states. The coupling between the driving electromagnetic field and qubits is characterized by the Rabi frequency. The decay of Rabi oscillations (transient nutations) provides direct information about the coherence time of the coupled field-qubit system. A detailed understanding of processes destroying the quantum coherence is of central importance for quantum computation. In EPR experiments, the decay of Rabi oscillations may results from the intrinsic decoherence in spin systems induced by couplings to the environment as well as from extrinsic decoherence induced by fluctuations and inhomogeneities of external magnetic or microwave fields [3,4]. Spatial inhomogeneities or fluctuations of the driving field can be a source of the so-called driven decoherence of the Rabi oscillations [4]. In particular, the effect of inhomogeneity in the driving amplitude over a spin ensemble increases with increasing this amplitude (the Rabi frequency) and can significantly shorten the decay of the Rabi oscillations. Changing the structure of the microwave field in a cavity or decreasing the volume of bulk samples not always can be used to overcome spatial driving-field inhomogeneities.

Decoherence arising from an inhomogeneity of polarizing magnetic field (i.e. from variations in the local magnetic field acting on each spin) can be reversed by a spin-echo technique [8]. Similarly, so-called rotary echoes permit refocusing the Rabi oscillations dephased by inhomogeneity in the driving field amplitude [9]. The rotary echo signals are generated by periodic pulse changes of the driving field phase by $\pi$ [9,10]. The rotary echo can also be formed by pulse variations in the resonant frequency of spins using the Zeeman effect [11]. However, complicated pulse sequences of the exciting fields are required in this technique.

On the other hand, a double dressing of spin qubits by a bichromatic field containing a strong transverse microwave (MW) and weak longitudinal radiofrequency (RF) components can also significantly extend their observed coherence times. Increasing in the decay time of Rabi oscillations between the doubly dressed states in a comparison with that between the singly dressed states was observed for spin ensembles in EPR [12-15] and NMR [16], when the radio frequency was in resonance with the Rabi frequency in the MW field. This secondary resonance used for the second dressing of spin states is sometimes termed rotary saturation [17], nutation resonance [14] or Rabi resonance [18,19]. It has been observed also for a single spin qubit [20]. However, the observed improvement in coherence times has not been studied in detail, though a completely different regime of interactions with the environment and a refocusing of the MW field inhomogeneity were mentioned as the possible origin of the observed improvement. Such refocusing differs essentially from the spin-locking method in which other relaxation dynamics is used to prolong the spin coherence time. In particular, keeping a system in long-lived singlet states by spin-locking has been proposed to increase the overall lifetime of the system [21]. Recently, it has been shown [22] that the double dressing modifies the dephasing and dissipative processes. This theoretical approach describes the dissipative dynamics of spin qubits assuming that there are no spatial inhomogeneities or fluctuations in the amplitudes of the MW and RF fields over a spin ensemble. The obtained analytical results can be useful in studies of the coherent dynamics at spatial inhomogeneity in the amplitude of the MW field.

In the present paper, we theoretically demonstrate that the double dressing of electron spin states by the bichromatic MW and RF field can suppress the dephasing effect of inhomogeneity of the MW field across the sample and enhances the observed decay times of Rabi oscillations towards their values limited by the spin relaxation processes. Such suppression is illustrated for experimental conditions corresponding to the EPR experiments with the Rabi oscillations between the doubly dressed spin states of paramagnetic nitrogen centers in diamond. We find that the maximum suppression of the inhomogeneity-induced decoherence is achieved at the Rabi resonance and is feasible even when the RF phase is random. Particularities of different types of the MW field inhomogeneity described by the cosine, Gaussian and linear distributions are considered. The manifestation of suppression of the observed spin decoherence at the Rabi resonance in some published EPR experiments is discussed.

## 2. Decay of Rabi oscillations between doubly-dressed spin states at a homogeneous driving

In this section we briefly review the theory developed in previous studies on the Rabi oscillations in the doubly dressed spin qubit [14,22,23]. We consider a spin qubit in three fields: a strong MW one oriented along the $x$ axis of the laboratory frame together with weak RF and static magnetic fields, both directed along the $z$ axis (Fig.1a).

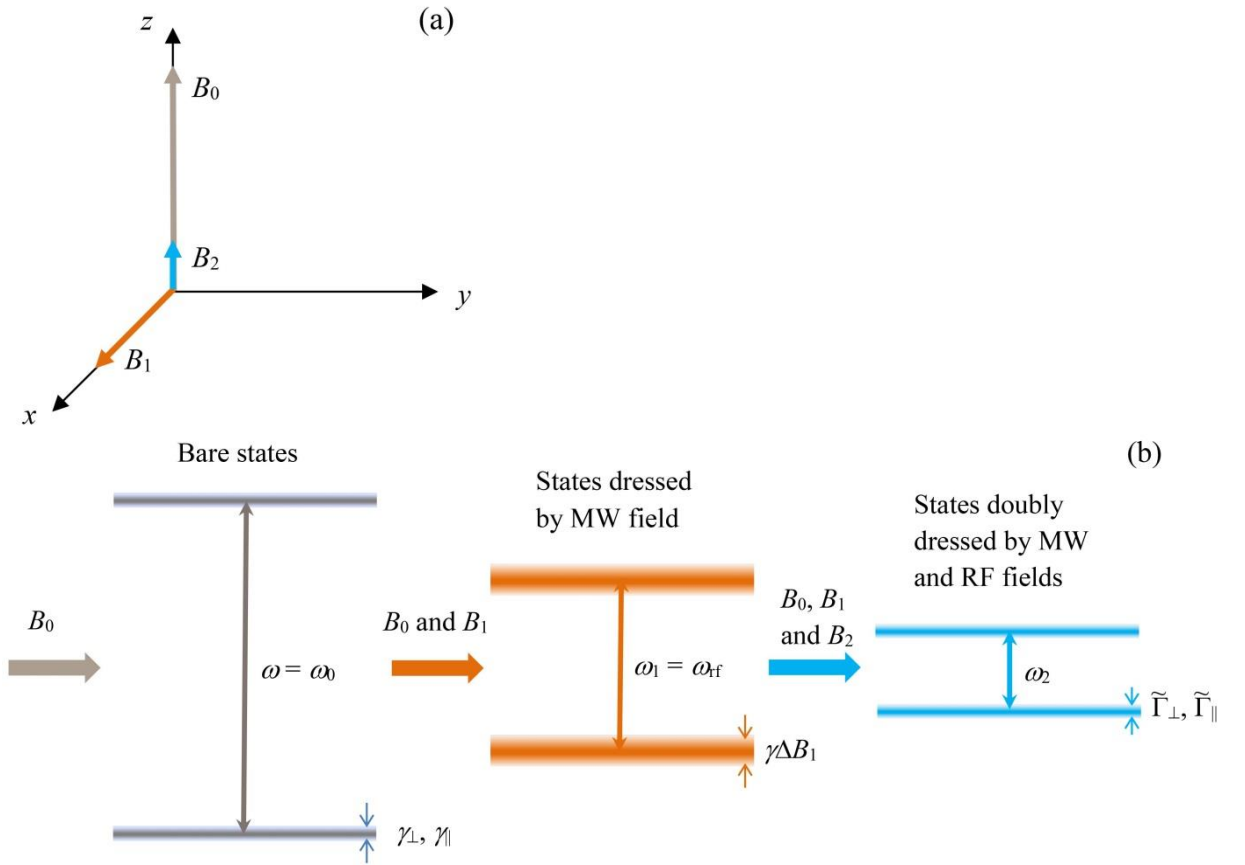

Fig. 1. (a) Field configuration in the laboratory frame. (b) Moving from the bare states in the homogeneous static magnetic field $B_0$ to the doubly dressed states using the bichromatic (MW + RF) field. The resonant ($\omega_{mw} = \omega_0$) MW field dresses each of initial spin states and splits into two states separated by the Rabi frequency $\omega_1 = \gamma B_1$. The RF field with $\omega_{rf} = \omega_1$

dresses each of dressed states and splits into new two states separated by the Rabi frequency $\omega_2 = \gamma B_2$. The inhomogeneity $\Delta B_1$ causes the distribution $\gamma \Delta B_1$ of the Rabi frequencies and the broadening of levels dressed by the MW field. The effect of the inhomogeneity may be reduced by making the RF field amplitude $B_2$ much larger than $\Delta B_1$.

The Hamiltonian of the qubit can be written as

$$H = H_0 + H_\perp(t) + H_\parallel(t), \qquad (1)$$

where $H_0 = \omega_0 s^z$ is the Hamiltonian of the Zeeman energy of the qubit in the static magnetic field $B_0$, $H_\perp(t) = \omega_1(s^+ + s^-)\cos\omega_{mw} t$ and $H_\parallel(t) = 2\omega_2 s^z \cos(\omega_{rf} t + \psi)$ are the Hamiltonians of the qubit interaction with linearly polarized MW and RF fields, respectively. Here $\omega_0 = \gamma B_0$, $\gamma$ is the electron gyromagnetic ratio, $\omega_1 = \gamma B_1$ and $\omega_2 = \gamma B_2$ are the Rabi frequencies, $B_1$ and $B_2$, $\omega_{mw}$ and $\omega_{rf}$ are the respective amplitudes and frequencies of the MW and RF fields, $\psi$ is the phase of the RF field, the MW phase being set to zero; $s^{\pm,z}$ are components of the spin operator, describing the state of the qubit and satisfying the commutation relations: $[s^+, s^-] = 2s^z$, $[s^z, s^\pm] = \pm s^\pm$. The rotating-wave approximation (RWA) is used for the interaction between the qubit and the MW field.

The dynamics of the qubit is described by the master equation for the density matrix $\rho$

$$i\hbar \frac{\partial \rho}{\partial t} = [H, \rho] + i\Lambda \rho \qquad (2)$$

(in the following we take $\hbar = 1$). The superoperator $\Lambda$ describing decay processes is defined as

$$\Lambda \rho = \frac{\gamma_{21}}{2} D[s^-]\rho + \frac{\gamma_{12}}{2} D[s^+]\rho + \frac{\eta}{2} D[s^z]\rho, \qquad (3)$$

where $D[O]\rho = 2O\rho O^+ - O^+ O \rho - \rho O^+ O$, $\gamma_{21}$ and $\gamma_{12}$ are the rates of the transitions from the excited state 2 of the qubit to its ground state 1 and vice versa, and $\eta$ is the dephasing rate.

Using the Krylov–Bogoliubov–Mitropolsky nonsecular perturbation theory on the parameter $\omega_2/\omega_{rf}$, from Eqs. (1) – (3) we can obtain the density matrix $\rho_{rot}$ [22] in the singly rotating frame (SRF), which rotates with frequency $\omega_{mw}$ around the $z$ axis of the laboratory frame. The MW field is much stronger than the RF field, $\omega_1 \gg \omega_2$. Moreover, we assume that $\Delta/\omega_{rf} \ll 1$, also $|\Omega - \omega_{rf}| \ll \Omega, \omega_{rf}$, and $\gamma_{21}, \gamma_{12}, \eta \ll \omega_{rf}, \Omega$, where $\Delta = \omega_0 - \omega_{mw}$, and

$\Omega = (\omega_1^2 + \Delta^2)^{1/2}$. In the SRF, the time-dependent absorption of the bichromatic field by the qubit can be written as

$$\upsilon = \frac{1}{2i}\left(\langle 1|\rho_{rot}|2\rangle - \langle 2|\rho_{rot}|1\rangle\right) = \frac{1}{2}N_{dd}\sin\xi(1-e^{-\tilde{\Gamma}_{\parallel}t})\sin(\omega_{rf}t+\psi) + \quad (4)$$

$$+\frac{1}{4}\left\{\sin\theta\sin^2\xi\left(\sin\omega_{rf}t+\sin(\omega_{rf}t+2\psi)\right) - 2\cos\theta\sin\xi\cos\xi\sin(\omega_{rf}t+\psi)\right\}e^{-\tilde{\Gamma}_{\parallel}t} +$$

$$+\frac{1}{8}\left\{\sin\theta\left[(\cos\xi+1)^2\sin(\omega_{rf}+\varepsilon)t + (\cos\xi-1)^2\sin(\omega_{rf}-\varepsilon)t - \right.\right.$$

$$-\sin^2\xi\left(\sin\left((\omega_{rf}+\varepsilon)t+2\psi\right) + \sin\left((\omega_{rf}-\varepsilon)t+2\psi\right)\right)\right] +$$

$$+2\cos\theta\sin\xi\left((\cos\xi+1)\sin\left((\omega_{rf}+\varepsilon)t+\psi\right) + (\cos\xi-1)\sin\left((\omega_{rf}-\varepsilon)t+\psi\right)\right)\right\}e^{-\tilde{\Gamma}_{\perp}t},$$

where $\sin\theta = \omega_1/\Omega$, $\cos\theta = \Delta/\Omega$, $\varepsilon = [\delta^2 + \omega_2^2\sin^2\theta]^{1/2}$, $\delta = \Omega - \omega_{rf} + \Delta_{BS}$, $\Delta_{BS} \approx \omega_2^2/4\omega_{rf}$, $\sin\xi = -\omega_2\sin\theta/\varepsilon$, $\cos\xi = \delta/\varepsilon$, $\tilde{\Gamma}_{\parallel} = \gamma_{\parallel} + (\gamma_{\perp} - \gamma_{\parallel})f(\theta,\xi)$, $\tilde{\Gamma}_{\perp} = \gamma_{\perp} - \frac{1}{2}(\gamma_{\perp} - \gamma_{\parallel})f(\theta,\xi)$, $f(\theta,\xi) = \sin^2\theta + (1-\frac{3}{2}\sin^2\theta)\sin^2\xi$, $\gamma_{\parallel} = \gamma_{12} + \gamma_{21}$, $\gamma_{\perp} = (\gamma_{12} + \gamma_{21} + \eta)/2$, $N_{dd} = -(\tilde{\Gamma}_{\downarrow} - \tilde{\Gamma}_{\uparrow})/\tilde{\Gamma}_{\parallel} = -(\gamma_{21} - \gamma_{12})\cos\theta\cos\xi/\tilde{\Gamma}_{\parallel}$ (if $\gamma_{12} \approx 0$, $\gamma_{21} \approx \gamma_{\parallel}$).

The longitudinal, $T_1$, and transverse (coherence), $T_2$, relaxation times of the bare qubit in the laboratory frame are represented by the respective relaxation rates as $\gamma_{\parallel} = 1/T_1$ and $\gamma_{\perp} = 1/T_2$.

Equation (4) describes the Rabi oscillations of the qubit in the bichromatic field and shows that in the SRF the Rabi oscillations occur at three frequencies, $\omega_{rf}$ and $\omega_{rf} \pm \varepsilon$. These oscillations result from the multiphoton transitions between the doubly dressed states of the qubit (Fig. 1b) [23]. The predicted frequencies of the Rabi oscillations were confirmed experimentally [13,14,20]. However, dephasing and relaxation of these states have not been studied experimentally in details until now. It follows from Eq. (4) that the doubly-dressed relaxation rates $\tilde{\Gamma}_{\parallel}$ and $\tilde{\Gamma}_{\perp}$ depend on a detuning of the MW frequency from the qubit resonant frequency as well as on a detuning $\delta$ between the radio frequency and $\omega_1$. At $\omega_2 = 0$ the expressions for $\tilde{\Gamma}_{\parallel}$ and $\tilde{\Gamma}_{\perp}$ reduce to those known for the singly dressed qubits [7].

## 3. Decay of Rabi oscillations between doubly-dressed spin states under an inhomogeneous MW driving

Besides dephasing and relaxation processes, inhomogeneities of the static magnetic field as well as the MW and RF fields across the sample can destroy the coherence of qubits increasing the decay of the Rabi oscillations. In particular, an inhomogeneity in the static magnetic field results in the distribution of resonant frequencies of qubits. It is known that near resonance ($\omega_{mw} \approx \omega_0$), the effect of inhomogeneity $\Delta B$ of the static magnetic field in the decay of Rabi oscillations may be reduced to an insignificant level by making $B_1$ much larger than $\Delta B$ [7, 24]. In this case the effective field acting on the spins in the rotating frame is almost the same as $B_1$, and the inhomogeneity $\Delta B$ is only a second-order effect. The elimination of inhomogeneity $\Delta B$ weakens with increasing the detuning $\Delta$ from the resonance. Note that in this method inhomogeneity $\Delta B$ is eliminated by a different process than in the spin-echo method which reverses dephasing of spin precession due to inhomogeneity $\Delta B$ [8]. However, at large $B_1$, inhomogeneity $\Delta B_1$ of the MW field becomes important and may result in the more rapid decay of Rabi oscillations than it is expected from relaxation processes. In the following, we assume that the static magnetic field is homogeneous and we consider the time evolution for qubits with a homogeneous EPR line in an inhomogeneous MW field. At the resonant MW field ($\omega_{mw} = \omega_0$), inhomogeneity $\Delta B_1$ may be expected to result in the decay of Rabi oscillations in a time of order of $1/\gamma\Delta B_1$.

At the double resonance in the bichromatic field ($\omega_{mw} = \omega_0$, $\omega_{rf} = \omega_1$), the effect of the inhomogeneity $\Delta B_1$ may be reduced by making the RF field amplitude $B_2$ much larger than $\Delta B_1$. In this case in the doubly rotating frame, which rotates with frequency $\omega_{rf}$ around the singly rotating frame, the homogeneous field $B_1$ is exactly canceled by the fictitious field $\omega_{rf}/\gamma$. In the presence of inhomogeneity $\Delta B_1$ the cancelation is incomplete, as illustrated in Fig. 2a. However, if $B_2 \gg \Delta B_1$, the effective field $\varepsilon/\gamma$ depends only slightly on the value of $\Delta B_1$ ($\varepsilon/\gamma \approx B_2$). Far from resonance ($\omega_{rf} \neq \omega_1$) the effective field $\varepsilon/\gamma$ has a much larger component along $B_1$ than at resonance $\omega_{rf} = \omega_1$. Consequently, the effect of inhomogeneity $\Delta B_1$ becomes stronger (Fig. 2b).

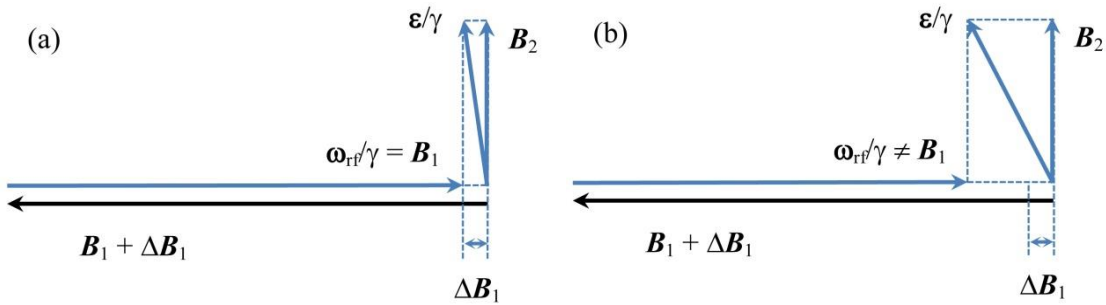

Fig. 2. The formation of effective field $\varepsilon/\gamma$ in the doubly rotating frame in the presence of an inhomogeneity $\Delta B_1$ in the MW field amplitude. (a) Spins are exactly at resonance $\omega_{rf} = \omega_1$. (b) Spins are far from resonance.

An inhomogeneity in the MW field amplitude gives rise to the distribution of the Rabi frequencies of qubits. The distribution of the Rabi frequencies across the sample is a complex function of the sample sizes and the structure of the MW field in the resonator. Often this distribution is unknown in detail. In some cases it can be approximated by the Gaussian function $g(\omega_1) = \frac{1}{\sqrt{2\pi}\sigma} \exp\left(-\frac{(\omega_1 - \omega_1^c)^2}{2\sigma^2}\right)$, where $\omega_1^c$ is the central Rabi frequency and $\sigma$ is the width of the distribution. We also study the Rabi oscillations for other types of an inhomogeneity in $B_1$ describing the distribution of $\omega_1$ by cosine and linear functions.

## 4. Results and discussion

In our calculations we use the relaxation parameters that are characteristic for such defects in solids as the isolated neutral substitutional nitrogen (P1 center) and the negatively charged nitrogen vacancy (NV) center in diamonds. The long-lived Rabi oscillations of these defects can be observed in the pulsed EPR even at room temperature. Their spin coherence time $T_2$ is changed from a few microseconds in nitrogen-rich diamond to a few milliseconds in ultrapure diamonds [5,25]. The NV centers are among the most promising candidates for solid-state qubits operating under ambient conditions [5].

Figure 3a displays the Rabi oscillations excited by the homogeneous MW field in the absorption EPR signal $\upsilon$ at the resonant MW driving, $\omega_{mw} = \omega_0$. The Rabi oscillations occur at the frequency $\omega_1$ and their decay is caused by the relaxation processes. The inhomogeneous MW field with $\gamma \Delta B_1 \gg \gamma_\perp, \gamma_\parallel$ gives rise to the distribution of the Rabi frequencies of qubits that result in the very fast decay of Rabi oscillations. The absorption signals $<\upsilon>$ averaged numerically over the distribution $g(\omega_1)$ are represented in Fig. 3a by the red and green lines for $g(\omega_1)$ approximated by the Gaussian and cosine functions, respectively.

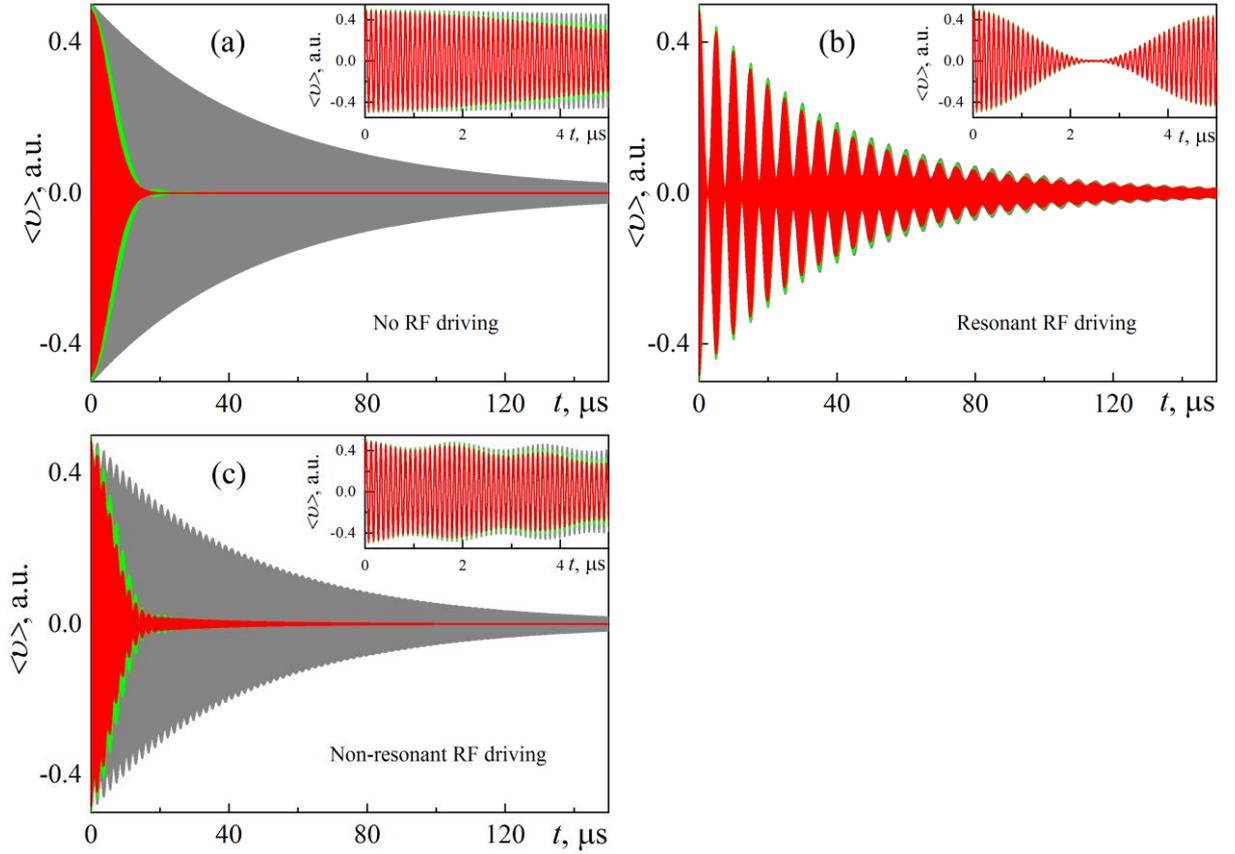

Fig. 3. The Rabi oscillations excited by the homogeneous (grey lines) and inhomogeneous (red and green lines) MW field without (a) and with the resonant (b) and non-resonant (c) RF driving. The parameters are $\omega_1^c = 2\pi\ 10.0$ MHz, $\Delta = 0$, $T_2 = 25$ μs, and $T_1 = 2$ ms. The resonant ($\omega_{rf} = \omega_1^c$) and non-resonant ($(\omega_1^c - \omega_{rf})/2\pi = 0.5$ MHz) RF field with the amplitude $\omega_2/2\pi = 0.2$ MHz is used. The distribution of $\omega_1$ is approximated by the Gaussian (red lines) and cosine (green lines) functions. For the Gaussian distribution $\sigma = 2\pi\ 0.028$ MHz. For the cosine distribution the relative change of the MW amplitude is about 0.006. The RF phase is random.

In order to demonstrate the suppression of decoherence in the Rabi oscillations using the second dressing of spin states by a longitudinal continuous RF field, we assume in our calculations that the static magnetic field and the RF field are homogeneous and that the decay rate of the Rabi oscillations is mainly due to the inhomogeneity of the MW field amplitude. Moreover, the RF field amplitude $B_2 \ll B_1$ and $B_2$ is much larger than the inhomogeneity $\Delta B_1$.

Figures 3b and 3c demonstrate the Rabi oscillations in the doubly resonant bichromatic field when the frequency of the RF field is equal or near to the Rabi frequency. The amplitude $B_2$ of the second driving is 50 times of magnitude weaker than the MW amplitude $B_1$ (i.e.

$\omega_2 \ll \omega_1$). The signals were calculated with and without taking into account effects of inhomogeneity in $\omega_1$. We assume that there are no random fluctuations in the amplitude of the MW field and average a large number of the oscillations assuming that the RF phase is random. In accordance with Eq. (4), at the bichromatic driving the Rabi oscillations occur at frequencies $\omega_{rf}$ and $\omega_{rf} \pm \varepsilon$. At the Rabi resonance ($\omega_{rf} = \omega_1^c$), $\varepsilon = \omega_2$ and the decay rate of the Rabi oscillations in the doubly resonant bichromatic field is clearly slower than the one in the monochromatic field. Figure 3c shows the oscillations at detuning the radio frequency from $\omega_1^c$. Besides an increase of the value of $\varepsilon$, detuning from the resonance $\omega_{rf} = \omega_1^c$ results in increasing the decay rate of the Rabi oscillations and at large values of the detuning the decay rate reaches its value observed in the MW field. There is no suppression of decoherence caused by the MW inhomogeneity.

The decay of Rabi oscillations as a function of detuning $\omega_1^c - \omega_{rf}$ for different types of inhomogeneity of the MW field is shown in Fig. 4. For the relative change of the MW amplitude of about 0.006 (i.e. 60 kHz), each type of such inhomogeneity results in the comparable decay of the Rabi oscillations in the MW field. At the bichromatic driving, the $\Delta B_1$-induced decay rate is much stronger sensitive to the detuning from the Rabi resonance than the decay rate in the homogeneous MW field. The decoherence from the all considered types of MW inhomogeneity is suppressed only at $\omega_{rf} = \omega_1^c$. The suppression is decreases with detuning $\omega_1^c - \omega_{rf}$ causing the faster decay. In particular, if there is a Gaussian distribution of the MW field amplitude of width $\sigma = 2\pi$ 0.028 MHz and a detuning of the radio frequency $(\omega_1^c - \omega_{rf})/2\pi =$ 0.5 MHz, the fastest decay time of the oscillations roughly decreases by a factor of 6.7 (Fig. 4c). At the homogeneous MW field and the same detuning, the change in the decay times is about 7% (Fig. 4a). Note that for the homogeneous MW field at the detuning $\delta/2\pi =$ 0.5 MHz and 0.9 MHz the decay rate is even smaller than the one is at the exact Rabi resonance (Fig. 4a). That is due to the different $\delta$-dependence of the decay rates and the amplitudes for the oscillations at $\omega_{rf}$ and $\omega_{rf} \pm \varepsilon$ [22]. For the symmetric distribution of $\omega_1$, the decay is the same for the different sign of detuning $\pm\delta$ (Fig. 4b and c). For the asymmetric distribution of $\omega_1$, the decay becomes dependent on the detuning sign (Fig. 4d). The presented results demonstrate that for the all considered types of the MW inhomogeneity the doubly resonant bichromatic field suppresses the dephasing effect of the inhomogeneity and increases the decay of the Rabi oscillations towards its value limited by the relaxation times.

Because the used RF field is weak ($\omega_2 \ll \omega_1$), the effect of its inhomogeneity is also much weaker than the one from the $B_1$-inhomogeneity. Therefore, it is neglected in our description. The inhomogeneities of the MW field cannot influence the homogeneity of the RF field because the relatively low-frequency continuous RF field is produced by the additional modulation coil. Unlike the longitudinal magnetic-field modulation, at the phase or frequency modulation of the MW field, the inhomogeneity in $B_1$ leads to the inhomogeneity in $B_2$.

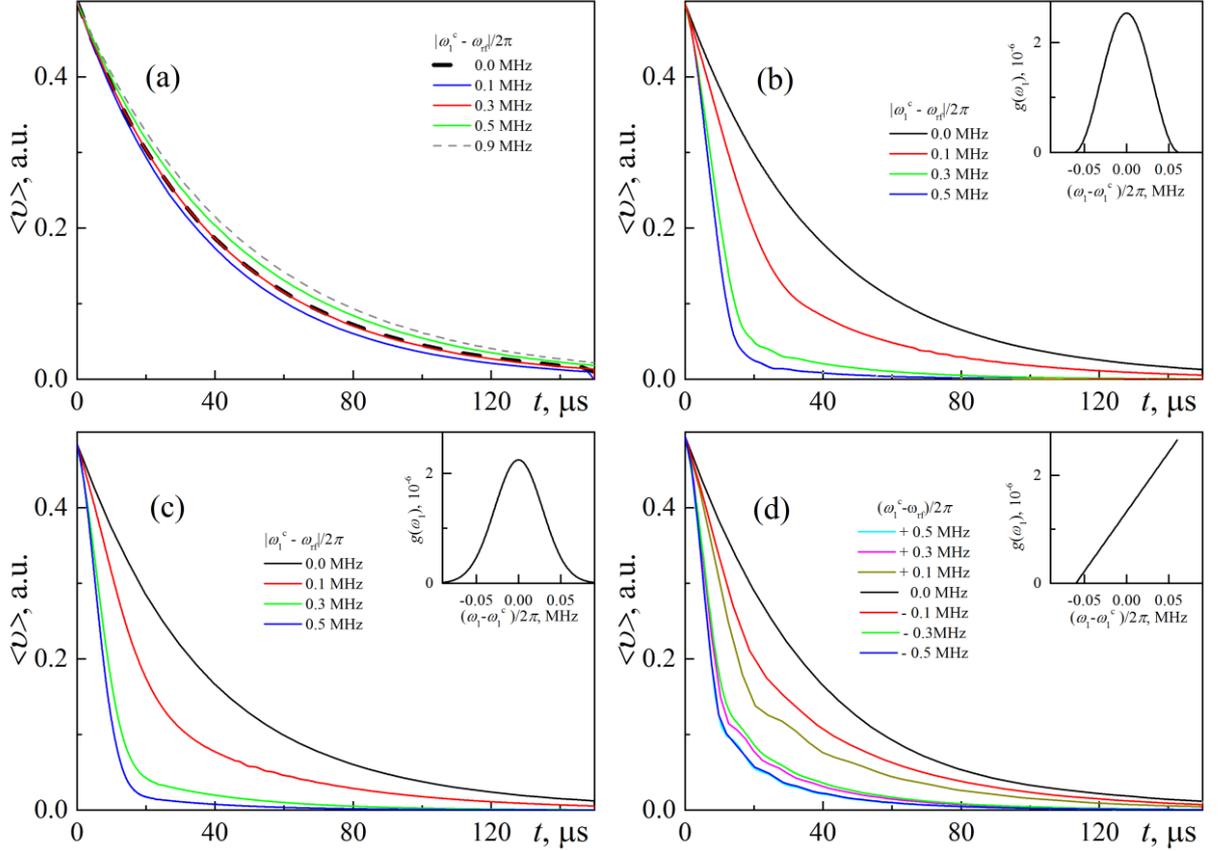

Fig. 4. The decay of Rabi oscillations excited by the bichromatic driving as a function of detuning $\omega_1^c - \omega_{rf}$ for different types of inhomogeneity of the MW field. The MW field is homogeneous (a). The distribution of $\omega_1$ is approximated by the cosine (b), Gaussian (c) and linear (d) functions and is shown in insets. Other parameters are the same as in Fig. 3.

As we found in Ref. [22], in the homogeneous MW and RF fields the amplitude of the Rabi oscillations at each frequency ($\omega_{rf}$ and $\omega_{rf} \pm \varepsilon$) is strongly dependent on the RF field phase [22]. This dependence can be used to realize conditions when the effect of inhomogeneity $\Delta B_1$ will be suppressed with the minimal distortion of the Rabi oscillations in the MW field. Fig. 5 illustrates the effect of the RF field phase in the inhomogeneous MW field. At the exact resonance $\omega_{rf} = \omega_1^c$ and the RF field phase $\psi = 0$, the signals at the frequencies $\omega_{rf} \pm \varepsilon$ are only due to the Bloch-Siegert effect in the RF field and they are much smaller than the signal at $\omega_{rf}$ [22]. For small values of $\omega_2$ ($\omega_2 \ll \omega_1$), the contribution of the oscillations at the frequencies $\omega_{rf} \pm \omega_2$ to the summarized signal is negligible. In this case the RF field suppresses the dumping of the Rabi oscillations due to inhomogeneity $\Delta B_1$ and the oscillations occur at the frequency $\omega_1$ without their modulation by the frequency $\omega_2$. Moreover, the decay of the Rabi oscillations is practically the same as the one observed at the homogeneous MW driving (Fig. 5a). At the exact resonance $\omega_{rf} = \omega_1^c$ and the RF phase $\psi = \pi/2$, the $\Delta B_1$-induced dumping of the Rabi oscillations is also suppressed (Fig. 5b). However, in this case the oscillations occur at frequencies $\omega_1 \pm \omega_2$ and their decay is something faster than the one at the homogeneous MW driving. That is because the decay rate $\tilde{\Gamma}_\perp$ of the oscillations at frequencies $\omega_1 \pm \omega_2$ is faster than the decay rate $\tilde{\Gamma}_\parallel$ of the oscillations at $\omega_1$. It is interesting that for the Gauss distribution of $\omega_1$ the double dressing recovers the decay rate to its value in the homogeneous MW field. For $\psi = 0$ and $\psi = \pi/2$, detuning from the resonance $\omega_{rf} = \omega_1^c$ destroys the suppression of decoherence (Fig. 5c and d).

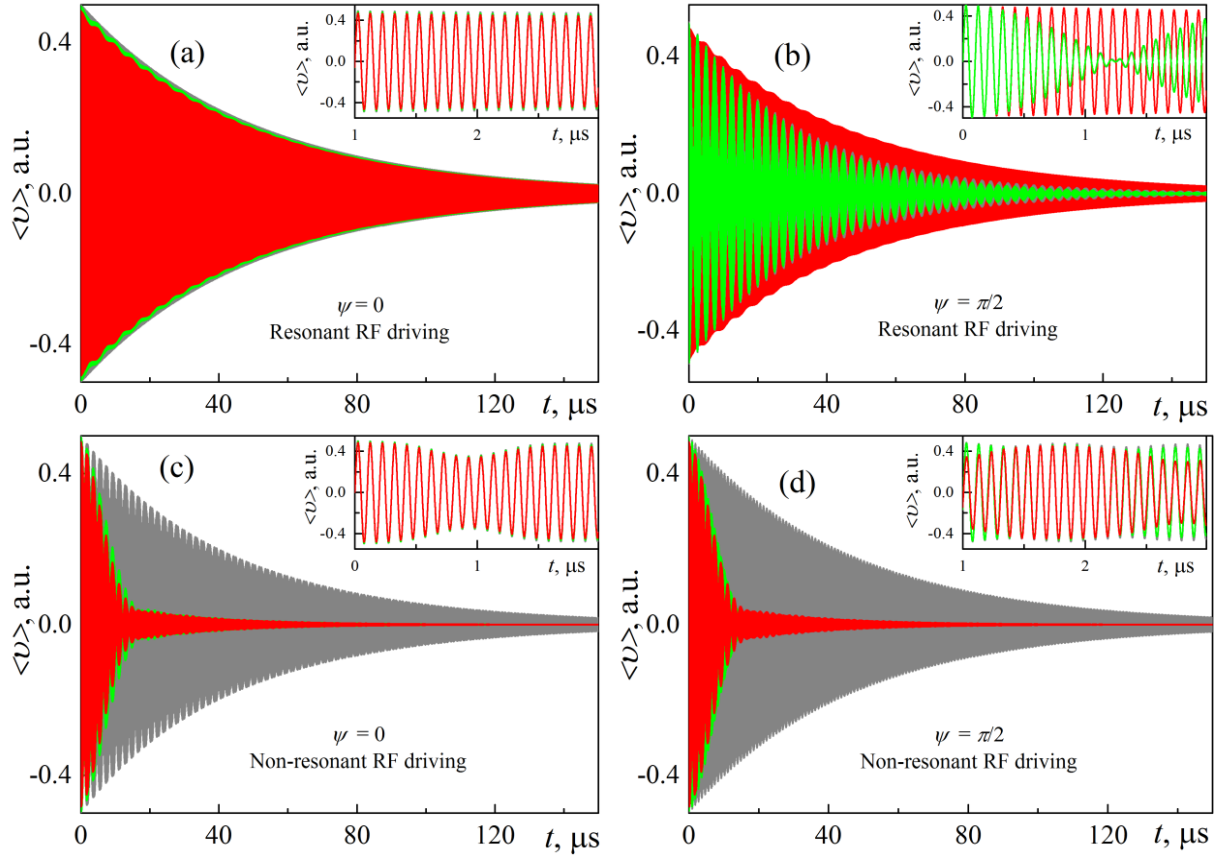

Fig. 5. The Rabi oscillations for two different values of the RF phase at the homogeneous

(grey lines) and inhomogeneous (red and green lines) MW field. The RF phase equals to 0 (a,c) and $\pi/2$ (b,d). Other parameters are the same as in Fig. 3.

To analyze decay kinetics, Fig. 6 shows in the logarithmic scale the decay of the Rabi oscillations at the Gaussian distribution of the MW field amplitude without and with the resonant and non-resonant RF driving. If no RF driving, the inhomogeneous MW field results in the fast decay which is well described by the Gaussian envelope $\exp(-a^2t^2/2)$ with $1/a = 5.5$ μs (line 1). Note that in the similar situation such decay was roughly approximated by a polynomial-type dependence [3,26]. The resonant RF driving suppresses decoherence from the MW inhomogeneity and the exponential decay with $1/b = 48.8$ μs occurs (line 2). This decay time is close to the value of 49.4 μs caused by the relaxation processes in the homogeneous MW field. At the non-resonant RF driving, the decay of the Rabi oscillations has a complex form (line 3). Its well-separated fast and slow components are approximated by $\exp(-a^2t^2/2)$ with $1/a = 6.6$ μs (the parabolic part of line 3) and $\exp(-bt)$ with $1/b = 27$ μs, respectively.

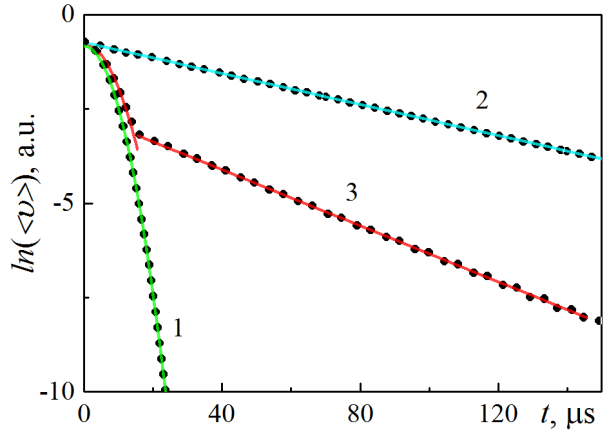

Fig. 6. The decay of the Rabi oscillations at the Gaussian distribution with $\sigma = 2\pi\ 0.028$ MHz of the MW field amplitude without (1) and with the resonant (2) and non-resonant (3) RF driving with the amplitude $\omega_2/2\pi = 0.2$ MHz and the phase $\psi = 0$. At the non-resonant RF driving $(\omega_1^c - \omega_{rf})/2\pi = 0.5$ MHz. Other parameters are $\omega_1^c = 2\pi\ 10.0$ MHz, $\Delta = 0$, $T_2 = 25$ μs, and $T_1 = 2$ ms. The straight and parabolic lines show the fit to the exponential and Gaussian functions, respectively.

In our paper the Rabi oscillations during the long MW pulse are considered. Therefore, it is difficult to compare directly this case with the existing sequences used to correct the MW field inhomogeneity [27-29]. However, we can estimate fidelity of the spin dynamics comparing the Rabi oscillations in the inhomogeneous MW and resonant RF fields with those in the homogeneous MW field (without the RF field). The last oscillations can be used as an ideal spin rotation. The four panels show the Rabi oscillations after 0, 50, 100, 150 μs and demonstrate the same frequency at the single MW and bichromatic driving even after 150 μs (Fig. 7). The decay in the homogeneous MW field is dictated purely by the relaxation processes. The decay in the inhomogeneous MW and resonant RF fields is slightly faster due to the residual rotation angle error. Our analytical and numerical calculations give fidelity of 0.988.

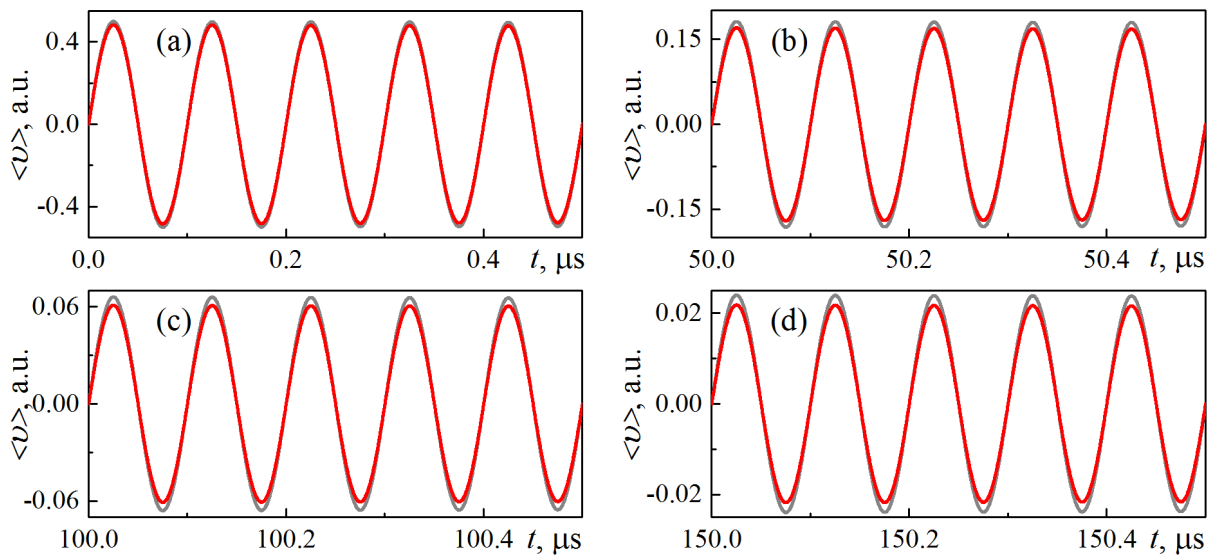

Fig. 7. The Rabi oscillation in the homogeneous MW field (grey lines) and in the inhomogeneous MW and resonant RF fields (red lines). Other parameters are $\omega_1^c = 2\pi\, 10.0$ MHz, $\Delta = 0$, $T_2 = 25$ μs, and $T_1 = 2$ ms, $\omega_2/2\pi = 0.2$ MHz, $\psi = 0$, $\sigma = 2\pi\, 0.028$ MHz.

As it was mentioned in Introduction, increasing in the decay time of Rabi oscillations was observed for spin ensembles in the pulsed EPR [13-15,30] and NMR [16] experiments with the double driving using the RF modulation of the magnetic field. However, the theoretical description of these experimental results was mainly oriented on the coherent dynamics of the doubly dressed states without the quantitative examination of decoherence. Our approach gives the opportunity to describe these results taking into account the modified relaxation between the doubly dressed states as well as the dephasing due to $B_1$ inhomogeneity. As an example, Fig. 8 depicts the simulation of the Rabi oscillations in the conditions similar to those in the EPR experiments with the bichromatic driving of P1 centers in a single diamond crystal [15,30]. There is a good agreement between the calculated and observed signals. Our calculations explain an increase of the spin coherence time almost ten times observed in these experiments via the decay of the Rabi oscillations. The resonant character of the improvement in the observed coherence time is a manifestation of the Rabi resonance. For the $E_1'$ centers in crystalline quartz such increase in the spin coherence time was about four times because the weaker $B_1$ inhomogeneity was in a volume of the smaller used sample [13,14].

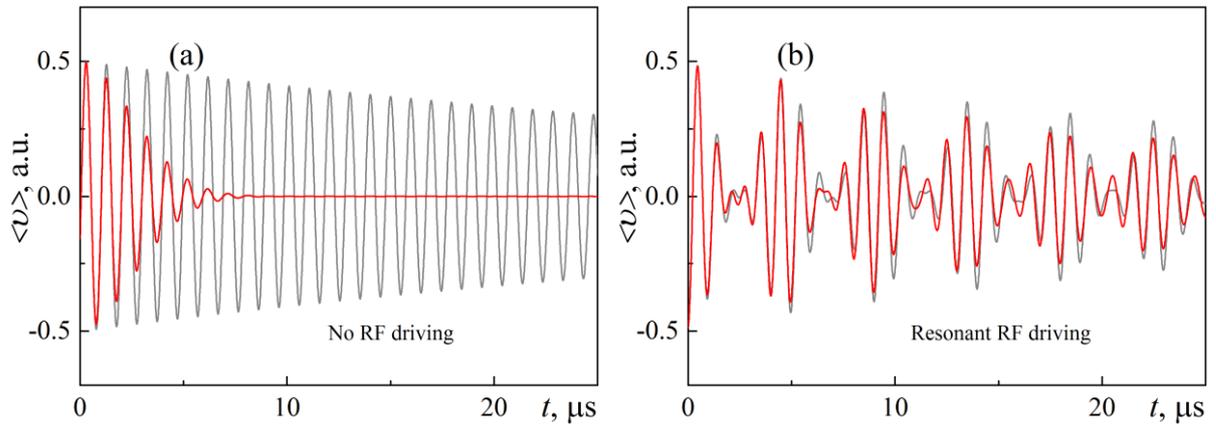

Fig. 8. The Rabi oscillations excited by the homogeneous (grey lines) and inhomogeneous (red and green lines) MW field without (a) and with the resonant (b) RF driving. The parameters are $\omega_1^c = \omega_{rf} = 2\pi$ 1.0 MHz, $\Delta = 0$, $\omega_2 = 2\pi$ 0.24 MHz, $T_2 = 25$ μs, and $T_1 = 2$ ms. The distribution of $\omega_1$ is approximated by the Gaussian function with $\sigma = 2\pi$ 0.05 MHz.

In some EPR experiments it is difficult to identify the source of the decay of Rabi oscillations and to separate the contributions of spin decoherence caused by relaxation and the inhomogeneous driving [3,4]. Our results demonstrate that the use of an additional longitudinal continuous RF field during the excitation of Rabi oscillations is useful for such separation.

In the considered approach for overcoming an inhomogeneity in the driving MW amplitude we apply an additional longitudinal continuous RF field. The same effective Hamiltonian can be generated by a weak additional transverse MW driving (see Appendix A). In this case the difference between two MW frequencies matches the Rabi frequency in the strong MW field. In order to avoid the use of an extra MW source, it is simpler to apply the weak RF field. However, this type of bichromatic transverse technique may be experimentally easier to implement in certain lower-field EPR applications and for certain sample geometries.

Recently, the amplitude [31] or phase [32] modulation of the transverse MW driving field was used to dress doubly spin qubits. An order of magnitude improvement in the coherence time for NV centers in diamond was experimentally and theoretically demonstrating with these modulation schemes. It was shown that these dynamical decoupling schemes protect against slow fluctuations in the MW driving amplitude. The fluctuation of MW field amplitude is much slower than the relaxation time $T_2$ in the scheme with a second-order driving field and it is much slower than $T_1$ for high-order driving fields. Both spatial inhomogeneities and slow fluctuations in the MW amplitude shorten the decay of the Rabi oscillations. Though the theoretical description of spatial and time changes in the MW field amplitude is different, the decoherence caused by both these sources can be refocused using a second weak driving to realize the Rabi resonance.

## 5. Conclusions

We have demonstrated for an electron spin system that the combination of the strong resonant MW excitation with a weak longitudinal continuous RF driving suppresses decoherence of the Rabi oscillations due to the MW field inhomogeneity. For the Gaussian, cosine and linear distributions of the microwave amplitude, it was shown that the maximum suppression is achieved when the radio frequency matches the Rabi frequency in the MW field. The resonant RF driving leads to the exponential decay of the oscillations which is free from any influence of the MW field inhomogeneity and causes only by the relaxation processes. At the non-resonant RF driving, the decay has a complex form with the fast (Gaussian) and slow (exponential) components. The fast component is mainly due to the MW amplitude distribution. The obtained results allowed us to interpret increasing in the decay time of Rabi oscillations observed previously for spin ensembles in the pulsed EPR experiments with the bichromatic driving. The dependence of the decay of the Rabi oscillations on the RF frequency was found to be strongly different in the inhomogeneous and homogeneous MW fields. The high sensitivity of the decay in Rabi oscillations to the detuning from the Rabi resonance at the MW field inhomogeneity open new possibilities for separating the contributions of relaxation mechanisms and the inhomogeneous driving in spin decoherence. Such discrimination of intrinsic and extrinsic decoherence of spin qubits can be useful in EPR and NMR experiments. The continuous RF modulation of the magnetic field is easier to realize experimentally than complicated pulse MW sequences. This approach can be combined with various quantum information processing tasks, eliminating the inhomogeneity-induced decoherence.

**Acknowledgment**

The work was supported by Belarusian Republican Foundation for Fundamental Research and by State Program of Scientific Investigations "Physical material science, new materials and technologies", 2016-2020.

**Appendix A. Spin qubit driven by two transverse MW fields**

We consider a spin qubit with the ground $|1\rangle$ and excited $|2\rangle$ states driven by two MW fields oriented along the $x$ axis of the laboratory frame. The Hamiltonian of the qubit can be written as

$$H = H_0 + H_1(t) + H_2(t), \tag{A1}$$

where $H_0 = \omega_0 s^z$ is the Hamiltonian of the qubit in the static magnetic field, $H_1(t) = \frac{\omega_1}{2}(s^+ e^{-i\omega t} + h.c.)$ and $H_2(t) = \frac{\omega_2}{2}(s^+ e^{-i\varpi t} + h.c.)$ are the Hamiltonians of the qubit interaction with strong and weak MW fields with frequencies $\omega$ and $\varpi$, respectively; $\omega_1$ and $\omega_2$ denote the respective interaction constants, and $s^{\pm,z}$ are components of the spin operator. We assumed that $\omega_1 \ll \omega$, $\omega_2 \ll \varpi$ and use the RWA for the interaction between the qubit and the MW fields. We also assume that $\omega_1 \gg \omega_2$, $\omega_0 \approx \omega$ and $\omega - \varpi = \omega_1 + \delta$, where $\delta \ll \omega_1$. The dynamics of the qubit is described by the corresponding master equation for the density matrix $\rho$ (see Eqs. (2) and (3)).

After three canonical transformations: $\rho' = u_3^+ u_2^+ u_1^+ \rho u_1 u_2 u_3$, where $u_1 = \exp(-i\omega t s^z)$, $u_2 = \exp(-i\pi s^y / 2)$ and $u_3 = \exp(-i\omega_1 t s^z)$, the master equation is transformed into $i\partial \rho'/\partial t = [H', \rho'] + i\Lambda' \rho'$. Here $H' = \frac{\omega_2}{4}\left[\left(e^{i\delta t}(e^{i2\omega_1 t} - e^{-i2\delta t})s^+ + h.c.\right) + 2s^z(e^{i\delta t}e^{i\omega_1 t} + h.c.)\right]$,

$\Lambda' \rho' = \frac{\gamma_\perp}{4} D[s^-]\rho' + \frac{\gamma_\perp}{4} D[s^+]\rho' + \frac{\gamma_\parallel}{2} D[s^z]\rho'$, $\gamma_\perp = (\gamma_{12} + \gamma_{21} + \eta)/2$, $\gamma_\parallel = \gamma_{12} + \gamma_{21}$.

Rapidly oscillating ($e^{\pm i\omega_1 t}$, $e^{\pm i 2\omega_1 t}$) terms in the Hamiltonian $H'$ can be eliminated by the Krylov–Bogoliubov–Mitropolsky method [33]. Assuming that $\omega_2/\omega_1 \ll 1$ and averaging over the period $2\pi/\omega_1$, we obtain the following effective Hamiltonian $H_{eff}$ up to the second order in $\omega_2/\omega_1$: $H' \to H_{eff} = H_{eff}^{(1)} + H_{eff}^{(2)}$ with $H_{eff}^{(1)} = \langle H'(t)\rangle = -(\omega_2/4)\left(s^+ e^{-i\delta t} + h.c.\right)$ and $H_{eff}^{(2)} = (i/2)\langle [\int^t d\tau (H'(\tau) - \langle H'(\tau)\rangle), H'(t)]\rangle = \delta_{BS} s^z$. In the above, the symbol $<...>$ denotes time averaging: $\langle A(t)\rangle = \frac{1}{T}\int_0^T A(t)dt$, where $T = 2\pi/\omega_1$ and $\delta_{BS} \approx \omega_2^2/16\omega_1$ is the Bloch–Siegert-like frequency shift. After the canonical transformation $\rho' \to \rho'' = u_4^+ \rho' u_4$, $u_4 = e^{-i\delta t s^z}$, the equation for $\rho'$ is transformed into $i\partial \rho''/\partial t = [H'', \rho''] + i\Lambda' \rho''$ with $H'' = (\delta + \delta_{BS})s^z - (1/4)\omega_2(s^+ + s^-)$. The solution of this equation can be written as

$$\rho'' = \frac{1}{2} + \frac{1}{2}\left[\left(e^{-\Gamma_\perp t}(\cos^2\xi \cos\varepsilon t - i\cos\xi \sin\varepsilon t) + e^{-\Gamma_\parallel t}\sin^2\xi\right)s^+ + h.c.\right] + \quad (A2)$$

$$+ \sin\xi\cos\xi(e^{-\Gamma_\parallel t} - e^{-\Gamma_\perp t}\cos\varepsilon t)s^z.$$

Here $\varepsilon = \left[(\omega - \varpi - \omega_1 + \delta_{BS})^2 + (\omega_2/2)^2\right]^{1/2}$, $\cos\xi = (\omega - \varpi - \omega_1 + \delta_{BS})/\varepsilon$,

$\Gamma_\perp = \frac{1}{2}(\gamma_\perp + \gamma_\parallel) + \frac{1}{4}(\gamma_\perp - \gamma_\parallel)\sin^2\xi$, $\Gamma_\parallel = \gamma_\perp - \frac{1}{2}(\gamma_\perp - \gamma_\parallel)\sin^2\xi$. Using Eq. (A2), in the SRF, the absorption signal is given by

$$\upsilon = \frac{1}{2i}\left(\langle 1|\rho_{rot}|2\rangle - \langle 2|\rho_{rot}|1\rangle\right) = \frac{1}{4}e^{-\Gamma_\perp t}\cos\xi\left[(\cos\xi+1)\sin(\omega-\varpi+\varepsilon)t + (\cos\xi-1)\sin(\omega-\varpi-\varepsilon)t\right] +$$

$$+ \frac{1}{2}e^{-\Gamma_\parallel t}\sin^2\xi \sin(\omega-\varpi)t. \quad (A3)$$

Eqs. (4) and (A3) have the same mathematical structures, if $\psi = 0$, $\Delta = 0$, $\omega_{rf}$ and $\omega_2$ are replaced by $\omega - \varpi$ and $\omega_2/2$, respectively. It means that at $\omega - \varpi = \omega_1$ the weak transverse MW can be used to suppress spin decoherence in Rabi oscillations induced by the inhomogeneous strong MW field."

**Graphical Abstract**

The Rabi oscillations excited by the homogeneous (grey lines) and inhomogeneous (red lines) MW field

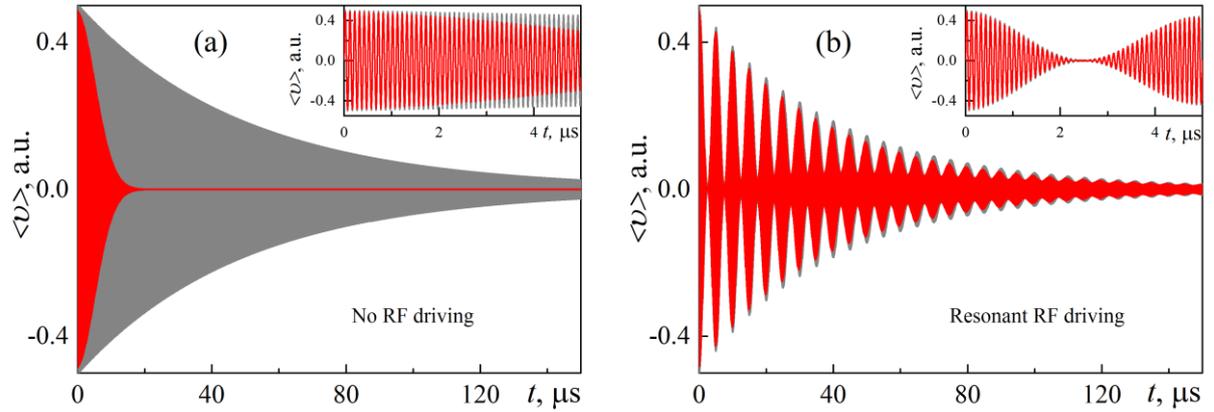

**Highlights**

- Suppression of decoherence in Rabi oscillations induced by inhomogeneous microwave field.
- Weak longitudinal radiofrequency field is used to suppress spin decoherence.
- The Gaussian, cosine and linear distributions of the microwave amplitude are analyzed.
- The maximum suppression of spin decoherence is achieved at Rabi resonance.
- A good agreement with the published experimental results is obtained.